\documentclass[prl,twocolumn,aps,a4paper,superscriptaddress]{revtex4-2}
\usepackage{graphicx}
\graphicspath{{./Figs/}}
\DeclareGraphicsExtensions{.pdf,.gif,.jpg}
\usepackage{dsfont}
\usepackage{bm}
\usepackage{amsmath,amssymb}
\usepackage{times}
\usepackage[colorlinks,linkcolor=blue,citecolor=blue,urlcolor=blue]{hyperref}
\usepackage[dvipsnames]{xcolor}
\usepackage{cleveref}

\newcommand{\mrho}{\bm{\mathit{\rho}}}

\newcommand{\be}{\begin{equation}}
\newcommand{\ee}{\end{equation}}
\newcommand{\bea}{\begin{eqnarray}}
\newcommand{\eea}{\end{eqnarray}}

\def\a{\alpha}
\def\b{\beta}

\def\d{\delta}
\def\D{\Delta}
\def\e{\epsilon}

\def\th{\theta}

\def\r{\rho}
\def\s{\sigma}
\def\S{\Sigma}

\def\w{\omega}

\def\ble{{\mathbf e}}

\def\blk{{\mathbf k}}

\def\blq{{\mathbf q}}

\def\bcallh{\mbox{\boldmath $h$}}

\def\bcallG{\mbox{\boldmath $\mathcal{G}$}}

\def\bcallV{\mbox{\boldmath $\mathcal{V}$}}

\def\bra{\langle}
\def\ket{\rangle}

\def\1op{\hat{\mathbbm{1}}}

\def\AA{\mathring{\mathrm{A}}}

\begin{document}

\title{Real-time $GW$: 
Toward an {\it ab initio} description of the ultrafast carrier and 
exciton dynamics \\in two-dimensional materials}

\author{E. Perfetto}
\affiliation{Dipartimento di Fisica, Universit\`{a} di Roma Tor Vergata,
Via della Ricerca Scientifica 1, 00133 Rome, Italy}
\affiliation{INFN, Sezione di Roma Tor Vergata, Via della Ricerca Scientifica 1, 00133 Rome, Italy}

\author{Y. Pavlyukh}
\affiliation{Dipartimento di Fisica, Universit\`{a} di Roma Tor Vergata,
Via della Ricerca Scientifica 1, 00133 Rome, Italy}

\author{G. Stefanucci}
\affiliation{Dipartimento di Fisica, Universit\`{a} di Roma Tor Vergata,
Via della Ricerca Scientifica 1, 00133 Rome, Italy}
\affiliation{INFN, Sezione di Roma Tor Vergata, Via della Ricerca Scientifica 1, 00133 Rome, Italy}

\begin{abstract}
We demonstrate the feasibility of the time-linear scaling formulation of the $GW$ method
[Phys. Rev. Lett. {\bf 124}, 076601 (2020)] for {\it ab initio} simulations of optically
driven two-dimensional materials. The time-dependent $GW$ equations are derived and solved
numerically in the basis of Bloch states.  We address carrier multiplication and
relaxation in photo-excited graphene and find deviations from the typical exponential
behavior predicted by the Markovian Boltzmann approach.  For resonantly pumped
semiconductor we discover a self-sustained screening cascade leading to the Mott
transition of coherent excitons.  Our results draw attention to the 
importance  of non-Markovian and dynamical screening effects in 
out-of-equilibrium phenomena.
\end{abstract}

\maketitle

In the last two decades the $GW$ approximation~\cite{PhysRev.139.A796} has emerged as a
successful and popular tool for describing at microscopic level electronic and optical
properties of quantum matter.  The merits of this method arise from the proper inclusion
of dynamical and non-local effects, leading to band structures and absorption spectra in
excellent agreement with experiments in a broad class of
materials~\cite{reining2018gw,10.3389/fchem.2019.00377}.

The nonequilibrium extension of the $GW$ approximation to address ultrafast phenomena has
been so far computationally prohibitive.  The real-time $GW$ method is based on the
numerical solution of the Kadanoff-Baym equations (KBE) for the one-particle Green's
function (GF). As the method scales cubically with the physical propagation
time~\cite{kadanoff1962quantum,svl-book,balzer2012nonequilibrium}, the $GW$--KBE have only
been applied to model
systems~\cite{mssvl.2009,mssvl.2008,pva.2010,pva.2009,PhysRevB.93.054303} and confined to
very short time scales~\cite{PhysRevB.100.235117}.  The time-scaling does not improve even
within the Generalized Kadanoff-Baym Ansatz (GKBA)~\cite{PhysRevB.34.6933}, and
simulations up to few hundreds of femtoseconds have been restricted to jellium-like
models~\cite{PhysRevLett.81.882,PhysRevLett.85.3508,PhysRevB.62.7179}.

A significant advance has been recently achieved with a time-linear scaling formulation of
the $GW$--GKBA approach~\cite{PhysRevLett.124.076601,PhysRevB.101.245101}.  The $GW$--GKBA
equations have been mapped onto a coupled system of ordinary differential equations (ODE)
for the one-particle density matrix and the equal-time 2-particle GF.  The ODE scheme
(also applicable to second-Born,
$T$-matrix~\cite{PhysRevLett.124.076601,PhysRevB.101.245101} and other correlated
methods~\cite{PhysRevLett.127.036402,PhysRevB.104.035124}) preserves the full
non-Markovian nature of the dynamics but it has been so far tested only in small finite
systems~\cite{PhysRevLett.124.076601,PhysRevB.101.245101,PhysRevB.104.035124}.

In this Letter we extend and solve numerically the $GW$-ODE scheme for spatially periodic
two-dimensional (2D) systems, thus opening the way to {\it ab-initio} real-time $GW$
simulations in material science.  We investigate two different materials to highlight
different aspects of the $GW$ method.  First we re-examine the problem of carrier
multiplication in photo-excited graphene~\cite{winzer2010carrier,PhysRevB.87.155429,brida2013ultrafast,plotzing2014experimental,
  tielrooij2013photoexcitation,PhysRevLett.109.166603,johannsen2015tunable}.  By
comparison with Boltzmann equation (BE) results, we show that the (so far neglected)
non-Markovian effects modify considerably the impact ionization dynamics.  The second
application concerns with the photogenerated screening in 2D semiconductors.  Pumping
resonantly with the exciton
energy~\cite{PhysRevLett.125.096401,man2021experimental,dong2021direct}, we find that
there is a critical excitation density above which the coherent exciton superfluid melts
abruptly ({\it coherent exciton Mott transition}) well before phonon-induced decoherence
takes places~\cite{nie2014ultrafast}. This is due to a self-sustained 
screening cascade, a phenomenon that can be captured only if the screened
electron-hole ($e$-$h$) attraction is properly updated during the evolution.

\begin{figure}[tbp]
    \includegraphics[width=0.45\textwidth]{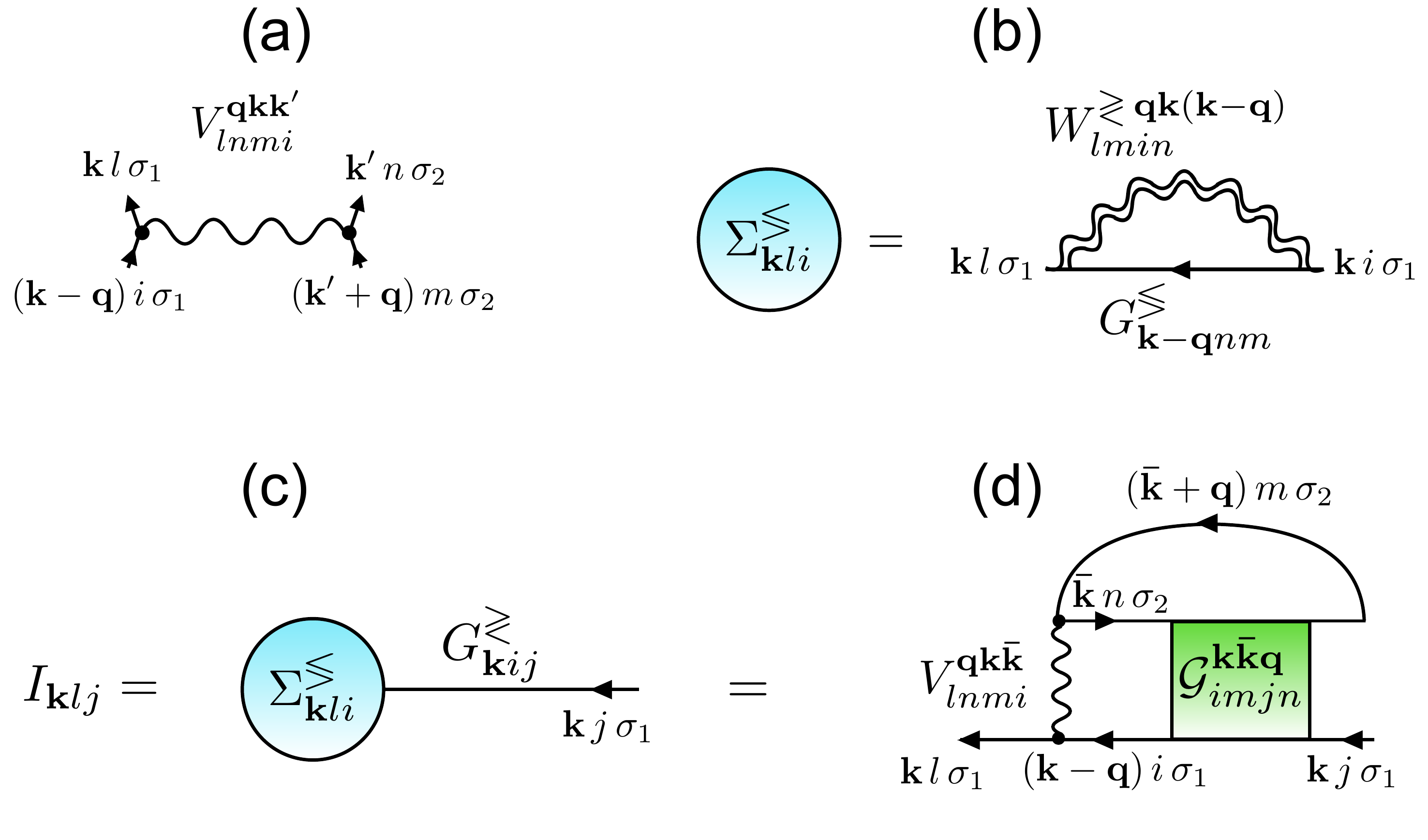}
    \caption{Diagrammatic representation of the Coulomb scattering amplitude (a), $GW$
      self-energy (b) and collision integral in terms of the self-energy (c), and
      two-particle GF (d). }
    \label{Diagrams_v3}
\end{figure}

{\it Real-time $GW$ formalism} -- We consider a periodic system with $N_{b}$ bands and denote by $V_{lnmi}^{\blq\blk \blk'}$
the scattering amplitude for two electrons in bands $m$ and $i$ with quasimomenta
$\blk'+\blq$ and $\blk-\blq$ to end up in the bands $n$ and $l$ with quasimomenta $\blk'$
and $\blk$ respectively, see Fig.~\ref{Diagrams_v3}(a). 
Let us introduce the spin-symmetric lesser and greater GFs
$G^{<}_{\blk ij}(t,t')=i \bra c^{\dag}_{\blk j \s}(t') c_{\blk i \s}(t) \ket $ and
$G^{>}_{\blk ij}(t,t')=-i \bra c_{\blk i \s}(t) c^{\dag}_{\blk j \s}(t') \ket $, where
$\hat{c}^{(\dagger)}_{\blk i \s}$ annihilates (creates) an electron with quasi-momentum
$\blk$ and spin $\s$ in band $i$.  The inclusion of spin-orbit and the generalization to
spin-dependent GFs is straightforward.  The goal of this work is to study the temporal
evolution of the one-particle density matrix $\r_{\blk ij}(t) \equiv -iG^{<}_{\blk
  ij}(t,t)$ in the $GW$ approximation.  By defining $\S_{\blk}(t,t')$ as the $GW$
self-energy, see Fig.~\ref{Diagrams_v3}(b), the equation of motion to solve is
\begin{align}
\frac{d\r_{\blk}(t)}{dt}&=-i[h_{\blk}(t),\r_{\blk}(t)]-I_{\blk}(t) -
I_{\blk}^{\dag}(t) , \label{eom} \\
I_{\blk}(t)&=\int  
dt'[\S_{\blk}^{>}(t,t')G_{\blk}^{<}(t',t)-\S_{\blk}^{<}(t,t')G_{\blk}^{>}(t',t)]
\label{coll}
\end{align}
where all quantities are $N_{b}\times N_{b}$ matrices in the band indices.  The
time-dependent single-particle Hamiltonian reads~\cite{PSMS.2015}
\begin{align}
h_{\blk ij}(t)&=\d_{ij}\e_{ \blk i} + P_{ \blk ij}(t) 
\nonumber \\
&+\sum_{\blk'mn}
(2V_{imnj}^{\bf{0} \blk\blk'}-V_{imjn}^{(\blk-\blk') \blk 
\blk'})\d\r_{\blk'nm}(t),
\label{hhf2}
\end{align}
where the $\e_{ \blk i}$ are the band-dispersions of a preliminary {\it equilibrium}
$GW$ calculation whereas $P_{ \blk ij}(t)$ describes the coupling of the electrons to an
external field. The last term in Eq.~(\ref{hhf2}) is the variation of the Hartree-Fock
potential due to the variation ($\d\r$) of the density matrix with respect to the
equilibrium value $\r^{\rm eq}_{\blk nm}=\d_{nm}f(\e_{ \blk n})$, with $f$ the
zero-temperature Fermi-Dirac distribution.

We implement the GKBA~\cite{PhysRevB.34.6933} to obtain a closed equation for $\r$, i.e.,
we express the lesser and greater GFs as $G_{\blk}^{\lessgtr}(t,t')=-G_{\blk}^{\rm
  R}(t,t')\rho_{\blk}^{\lessgtr}(t') +\rho_{\blk}^{\lessgtr}(t)G_{\blk}^{\rm A}(t,t')$,
where we have defined $\r_{\blk}^{<}=\r_{\blk}$ and $\r_{\blk}^{>}=\r_{\blk}-
\mathds{1}$. The retarded and advanced propagators are approximated at the quasi-particle
level, hence $G^{\rm
  R}_{\blk}(t,t')=-i\th(t-t')T\{e^{-i\int_{t'}^{t}d\bar{t}\,h_{\blk}(\bar{t})}\}$ ($T$
being the time-ordering operator) and $G^{\rm A}_{\blk}(t,t')=[G^{\rm
    R}_{\blk}(t',t)]^{\dag}$.  The bottleneck in solving Eq.~(\ref{eom}) is the numerical
calculation of the $GW$ self-energy $ \S^{\gtrless}_{\blk ij}(t,t')=i\sum_{\blq
  mn}W^{\gtrless \, \blq \blk (\blk-\blq) }_{imjn} (t,t') G^{\lessgtr}_{\blk-\blq
  nm}(t,t')$ since the screened interaction $W(t,t')$ obeys an RPA-like integral equation
for every $t$ and $t'$, making the overall numerical scaling cubic in the propagation
time.  Remarkably, however, such scaling can been reduced from cubic to linear through the
simultaneous propagation of the two-particle GF
$\mathcal{G}$~\cite{PhysRevLett.124.076601,PhysRevB.101.245101}.  The collision integral
$I(t)$ in Eq.~(\ref{coll}), see also Fig.~\ref{Diagrams_v3}(c), can equivalently be
written in terms of $\mathcal{G}$ as illustrated in Fig.~\ref{Diagrams_v3}(d):
\be I_{\blk
  lj}(t) = -i\sum_{\begin{subarray}{c} \bar{\blk} \blq \\ imn \end{subarray}}
\bm{\mathcal{G}}^{\blk \bar{\blk}\blq } _{\begin{subarray}{c} in \\ jm \end{subarray}} (t)
\bcallV^{(-\blq)\bar{\blk}\blk }_{\begin{subarray}{c} ni \\ ml
\end{subarray}} ,
 \label{coll3}
\ee
where $\bm{\mathcal{G}}^{\blk \blk' \blq} _{\begin{subarray}{c} lm \\ in \end{subarray}} =
\mathcal{G}_{lnim}^{\blk \blk' \blq}$ and $\bcallV^{\blq \blk \blk'}_{\begin{subarray}{c}
    lm \\ in \end{subarray}} = V_{lnmi}^{\blq \blk \blk'}$ are matrices (two-index
tensors) in the space of pairs of band indices~\cite{citesupptdgw}. Henceforth we use
boldface letters to denote matrices in this space.  Introducing also the matrices $
\bcallh_{\blk \blk' \, {\begin{subarray}{c} lm \\ in \end{subarray}}}(t) \equiv h_{\blk
  lm}(t)\d_{in} - h_{\blk'ni}\d_{lm}(t)$ and $\mrho^{\lessgtr}_{\blk \blk' \,
  {\begin{subarray}{c} lm \\ in \end{subarray}}}(t)\equiv\r^{\lessgtr}_{\blk lm}
(t)\r^{\gtrless}_{\blk' ni} (t)$ we obtain a compact equation of motion for the 2-particle
GF~\cite{citesupptdgw}:
\begin{widetext}
\begin{align}
i\frac{d}{dt}\bm{\mathcal{G}}^{\blk \bar{\blk}\blq}(t) &=
-2\left[ \mrho^{>}_{(\blk -\blq) \blk }(t) \bcallV^{\blq (\blk-\blq) 
(\bar{\blk}+\blq)} \mrho^{<}_{ \bar{\blk} (\bar{\blk} +\blq) }(t)- 
\mrho^{<}_{(\blk -\blq) \blk }(t) \bcallV^{\blq (\blk-\blq) 
(\bar{\blk}+\blq)} \mrho^{>}_{ \bar{\blk} (\bar{\blk} +\blq) }(t)     
\right]   \nonumber \\
&+ \bcallh_{(\blk -\blq) \blk} (t ) \bm{\mathcal{G}}^{\blk \bar{\blk}\blq}(t)
-\bm{\mathcal{G}}^{\blk \bar{\blk}\blq}(t) \bcallh_{ \bar{\blk} 
(\bar{\blk} +\blq) } (t ) \nonumber \\
&+ 2 \sum_{\blk'} \left[ \bm{\mathcal{G}}^{\blk \blk'\blq}(t) 
\bcallV^{(-\blq) \blk' (\bar{\blk}+\blq)} \mrho^{\D}_{ \bar{\blk} 
(\bar{\blk} +\blq) }(t) -
\mrho^{\D}_{ (\bar{\blk}-\blq)\blk }(t) \bcallV^{(-\blq) (\blk-\blq) 
\blk'}\bm{\mathcal{G}}^{\blk' \bar{\blk}\blq}(t) 
\right] \equiv \bm{\mathcal{I}}^{\blk \bar{\blk}\blq}(t).
\label{eomG2}
\end{align} 
\end{widetext}
In Eq.~(\ref{eomG2})
$\mrho^{\D}_{\blk \blk' }\equiv\mrho^{>}_{\blk \blk' }-\mrho^{<}_{\blk \blk'}$
and matrix multiplication between $N_{b}^{2}\times N_{b}^{2}$ 
matrices  is understood.

Equations~(\ref{eom},\ref{coll3},\ref{eomG2}) form a closed system of ODE that is
equivalent to the original $GW$--GKBA scheme.  Notice that in the $GW$--ODE scheme the
$GW$ self-energy $\S(t,t')$ is never evaluated.  The collision integral
$\bm{\mathcal{I}}(t)$ depends only on the {\it instantaneous} $\bcallG(t)$, $\mrho(t)$ and
$\bcallh(t)$.  The numerical scaling is linear in time, quartic with the number of $\blk$
points and sextic with the number of bands.  The second-Born (2B) approximation without
the second-order exchange contribution is recovered by neglecting the last line of
Eq.~(\ref{eomG2}).  If we instead set $N_{b}=1$ and choose $V_{1111}^{\blq\blk
  \blk'}=V^{\blq}$ depending only on the transferred momentum $\blq$,
Eqs.~(\ref{eom},\ref{coll3},\ref{eomG2}) reduce to the $GW$--ODE for
jellium~\cite{PhysRevLett.124.076601,PhysRevB.101.245101}.

We have implemented the $GW$--ODE scheme in 2D systems having a single valence ($i=1$) and
a single conduction ($i=2$) band (hence $N_{b}=2$). Accordingly, the equilibrium
one-particle density matrix $\r_{\blk 11}^{\rm eq}=1$, $\r_{\blk 22}^{\rm eq}=0$, and
$\r_{\blk 12}^{\rm eq}=0$.  We use $\r_{\blk}(0)=\r_{\blk}^{\rm eq}$ as initial condition.
To avoid double countings we also subtract from the right hand side of Eq.~(\ref{eomG2})
the contribution of initial correlations, already taken into account in the the dressing
the $GW$ band structures $\e_{\blk i}$.  Hence we modify the equation of motion for
$\bm{\mathcal{G}}$ according to $ i\frac{d}{dt}\bm{\mathcal{G}}^{\blk \bar{\blk}\blq}(t)=
\bm{\mathcal{I}}^{\blk \bar{\blk}\blq}(t)-\bm{\mathcal{I}}^{\blk \bar{\blk}\blq}(0) \, $
and set $\bm{\mathcal{G}}^{\blk \bar{\blk}\blq}(0)=0$.  Although the equilibrium state is
weakly correlated, the electron-electron interaction plays a crucial role in the
photo-excited dynamics, see below.  We solve numerically the $GW$--ODE equations for the
$GW$ and 2B approximations using the CHEERS code~\cite{PS-cheers}; we also provide
comparisons with results from the BE, i.e., the semiconductor Bloch 
equation~\cite{KIRA2006155} in the 2B-Markov approximation~\cite{citesupptdgw}.

{\it Carrier multiplication in photo-excited graphene} -- 
Due to its semimetallic nature pristine graphene has
scarce screening efficiency~\cite{PhysRevLett.77.3589,PhysRevB.76.115434}.  Moreover in
Ref.~\cite{PhysRevB.102.045121} it has been shown that second-order exchange effects are
negligible.  We therefore expect that 2B and $GW$ calculations give similar results and
that the comparison between $GW$ and BE well highlights the role of non-Markovian effects.

 \begin{figure}[tbp]
     \includegraphics[width=0.45\textwidth]{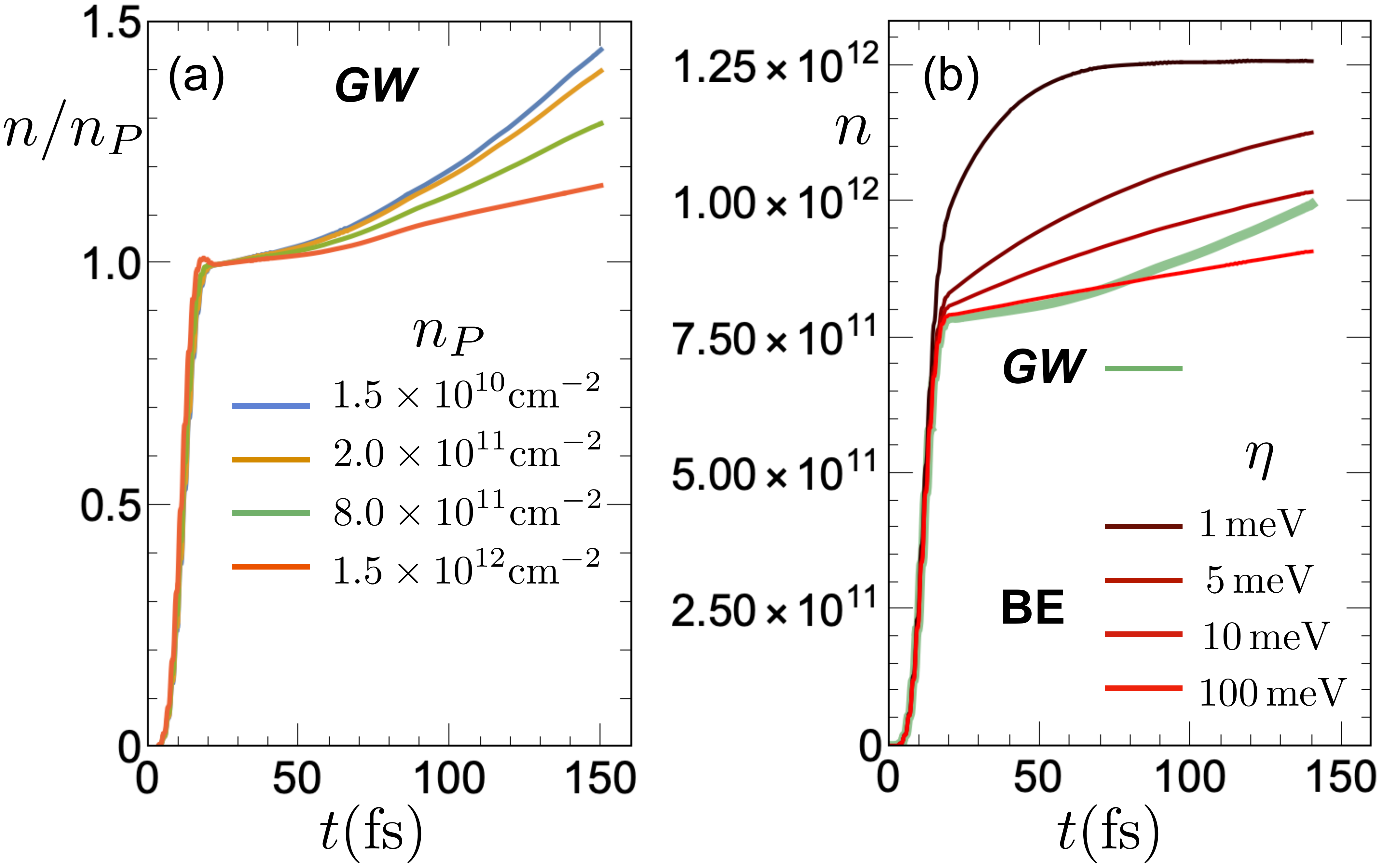}
  \caption{(a) Time-dependent normalized carrier density $n(t)/n_{P}$ in photo-excited
    graphene in the $GW$ approximation for different pump 
    intensities. (b) Carrier
    density $n(t)$ in the $GW$ and BE approach for different values of the broadening
    $\eta$; in all simulations we use the same pump intensity, giving
$n_{P}=8\times 10^{11}\mathrm{cm}^{-2}$ in the $GW$ case.  }
  \label{fig2}
  \end{figure}

Previous studies have shown that immediately after the photoexcitation, the electron
dynamics is dominated by the impact
ionization~~\cite{PhysRevB.76.155431,winzer2010carrier,plotzing2014experimental,PhysRevB.84.205406,PhysRevB.88.035430,PhysRevB.97.205411}.
This inter-band process promotes electrons from the valence to the conduction band at the
expense of energy loss by photo-excited hot carriers.  Due to the linearity of the Dirac
spectrum, carrier multiplication can occur mainly via {\it collinear
scattering}~\cite{PhysRevB.88.035430}. These scatterings, however, 
have a vanishingly small phase-space (and therefore 
become irrelevant) if the energy of the 
quasiparticles is exactly conserved~\cite{PhysRevB.88.035430}. This means that 
in the BE approach  an empirical energy--broadening $\eta$~\cite{citesupptdgw} must be introduced
to capture the effect.

%
%

For photoexcitations with photon energy $\lesssim 3$~eV graphene is well described by the
Dirac cone approximation~\cite{RevModPhys.81.109}, where conduction and valence bands have
linear dispersion $\e_{\blk 1,2}=\pm v_{F} k$, with $v_{F}$ the Fermi velocity and
$k=|\blk|$ a small momentum around the K (K') point of the first Brillouin zone.  In this
case the Coulomb integral has a simple expression~\cite{RevModPhys.81.109} $V_{lnmi}^{\blq
  \blk \blk'}=\frac{2\pi}{\e q}F_{il}(\theta_{\blk -\blq}-\theta_{\blk})
F_{mn}(\theta_{\blk' +\blq}-\theta_{\blk'})$, where
$F_{\a\b}(\theta)=\frac{1+(-1)^{\a+\b}e^{i\theta}}{2}$, with $\theta_{\blk}$ the polar
angle of the momentum $\blk$.  We take a dielectric constant $\e \approx 2.5$, originating
from a typical insulating substrate like SiO$_{2}$~\cite{doi:10.1143/JPSJ.75.074716}.

We consider graphene initially in the ground state and then driven out of equilibrium by a
pump field linearly polarized along a direction $\ble$ on the plane.  The explicit form of
the light-matter interaction term is~\cite{PhysRevB.67.165402} $P_{\blk
  ij}(t)=\d_{i1}\d_{j2}M E(t) \frac{k_{x}e_{y}-k_{y}e_{x}}{k}$, where the
$E(t)=\theta(1-|1-2t/T_{P}|)E\sin^{2}\left(\frac{\pi t}{T_{P}} \right)\sin (\w_{P}t)$ is
the pump envelope with duration $T_{P}=20$~fs and frequency $\w_{P}=1.5$~eV; the Rabi
frequency $M$ is varied in order to promote excitation densities in the range
$10^{10}-10^{12}$ carriers$/\mathrm{cm}^{2}$.  To improve convergence we have regularized
the bare interaction $1/q \to 1/(q+q_{c})$; in the simulations $q_{c}=0.01 ~ \AA^{-1}$ is
a small cutoff that can be understood as the Thomas-Fermi momentum ascribed to a small
unintentional doping~\cite{PhysRevB.75.205418}.  The EOM have been solved numerically by
simulating the carrier dynamics inside the K (K') valley up to a time $150$~fs. At times
$\gtrsim 200$~fs intervalley scattering and electron-phonon interactions (which are
neglected in our calculations) start to be
relevant~\cite{PhysRevB.84.205406,PhysRevB.88.035430} and our theory becomes less
accurate.  In Fig.~\ref{fig2}a we show the evolution of the carrier density in the
conduction band $n(t)=\frac{4}{A}\sum_{\blk}\r_{\blk 22}(t)$ during and after the
illumination, for different pump intensities.  The factor $4$ accounts for the spin and
valley degeneracy while $A=5.1\times 10^{-16}\mathrm{cm}^{2}$ is the unit-cell area of
graphene.  In order to illustrate the carrier multiplication effect as a function of the
pump intensity we plot $n(t)/n_{P}$, where $n_{P} \equiv n(T_{P})$ is the excited density
at the end of the pump.  The $GW$ simulation confirms the predicted behavior that the rate
of carrier multiplication decreases by increasing the carrier
density~\cite{winzer2010carrier}.  This is due to a Pauli blocking effect that reduces the
phase-space for impact ionization. Notice that no parameters (like $\eta$ in the BE
approach) appear in the $GW$-ODE scheme.

In Fig.~\ref{fig2}b we compare the $GW$ result to the BE outcome for different values of
the broadening $\eta$. In all simulations we use the same pump intensity, giving
$n_{P}=8\times 10^{11}\mathrm{cm}^{-2}$ in the $GW$ case.  We see that the carrier
multiplication effect predicted by the BE approach depends strongly on the chosen
broadening. During illumination, energy is not conserved and therefore the smaller $\eta$
is, the less accurate is the description of the early transient dynamics for $t<T_{P}$.
This explains why at the beginning the BE curve obtained with the large value
$\eta=0.1$~eV is the closest to the $GW$ one.  At larger times $t \gtrsim 60$~fs, however
the two curves depart from each other. In particular the $GW$ evolution does not follow
(at least within this temporal window) the typical exponential saturation behavior of the
BE, characterized by a downward concavity for $t>T_{P}$.  This qualitative difference is
due to non Markovian effects as in this case the 2B results (not shown) are very close to
the $GW$ one.

\begin{figure}[tbp]
     \includegraphics[width=0.45\textwidth]{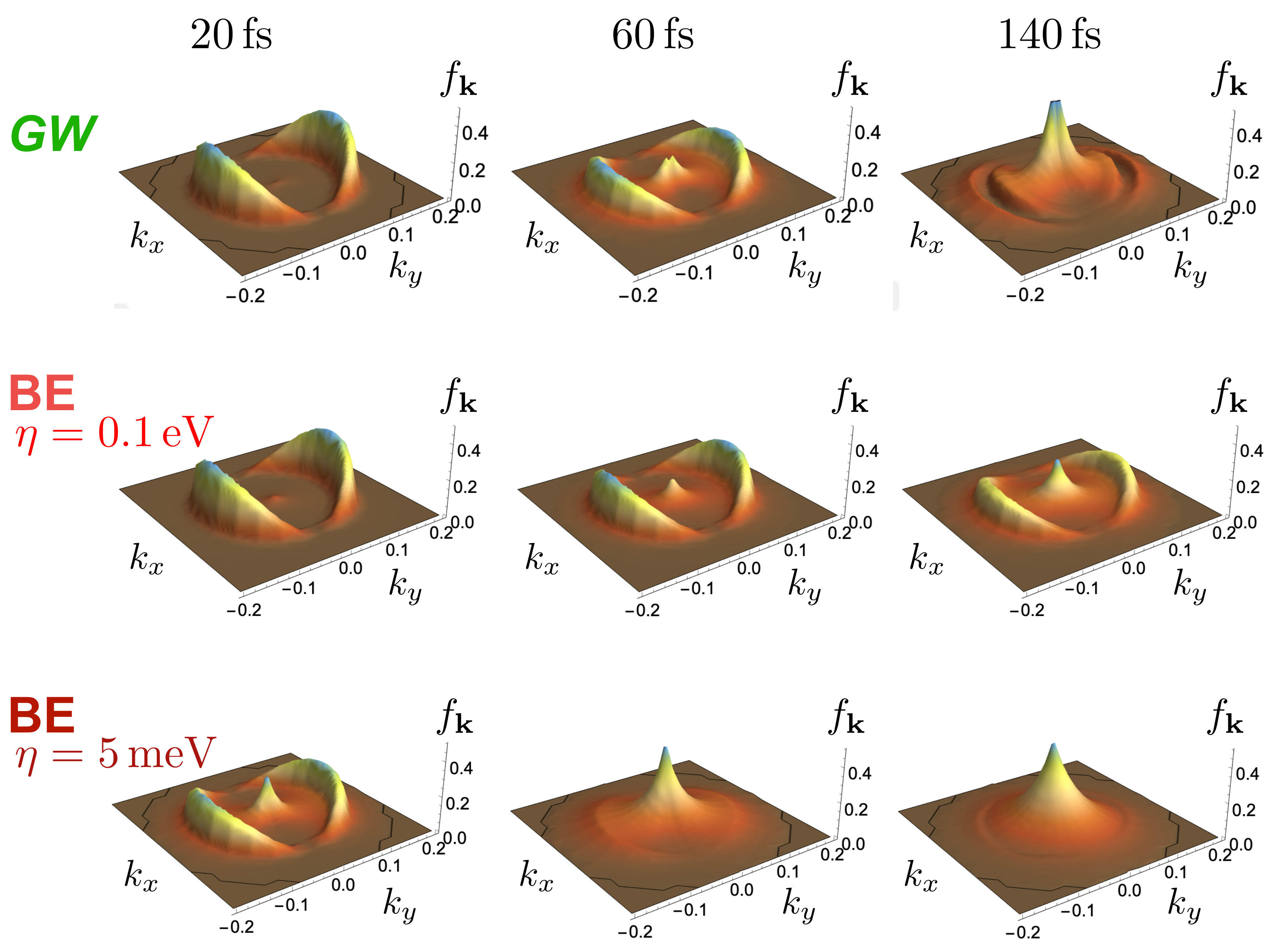}
  \caption{Snaphots of the carrier distribution function $f_{\blk}$ in photo-excited
    graphene in $GW$ (1st row), BE for $\eta=100$~meV (2nd row) and BE $\eta=5$~meV (3rd
    row).  Here the pump field is polarized along the $x$ axis. 
    Momenta $k_{x}$ and $k_{y}$ are in $ \AA^{-1}$. }
  \label{fig3}
\end{figure} 

In Fig.~\ref{fig3} we compare the evolution of the momentum-resolved occupations
$f_{\blk}(t)=\r_{\blk 22}(t)$ in different approaches.  We clearly see that the carrier
population is initially highly anisotropic due to the liner polarization of the
pulse~\cite{PhysRevB.84.205406}.  As already observed $GW$ agrees well with BE for
$\eta=0.1$~eV up to time $t\approx 60$~fs, i.e., when the distribution is still
anisotropic.  At this time a substantial portion of the initial hot electrons have already
migrated towards the Dirac point due to interparticle scattering.  At larger times $GW$
predicts a rapid thermalization while in BE the same process is much slower.  The BE
results are strongly affected by the value of $\eta$.  At smaller $\eta=5$~meV the
themalization is very fast: there is a sizable charge redistribution already during
illumination, and at $t\approx 60$~fs the distribution is essentially isotropic.

A crucial feature of the real-time $GW$ method is the updating of the screened interaction
during the time evolution.  We highlight this effect in a prototype 2D semiconductor
hosting bound excitons inside the gap, and study the dynamics activated by pumping in
resonance with the lowest excitonic energy.  A fluid of coherent excitons is then
formed~\cite{Schmitt-Rink_PhysRevB.37.941}, characterized by long-lived coherent
oscillations of the macroscopic
polarization~\cite{kuklinski1990,Glutsch_PhysRevB.45.5857,littlewood1996,Ostreich_1993,Hannewald-Bechstedt_2000,GLUTSCH1992,PSMS.2019}.
We here address the relaxation dynamics of the macroscopic polarization due to excited
state screening.  Let us model a direct-gap 2D semiconductor with band dispersions
$\e_{\blk 1,2}=\pm \e_{g}/2 \pm k^{2}/2m$, where $\e_{g}$ is the bandgap and $m$ the
effective mass of electrons and holes.  In semiconductors the Coulomb integrals that do
not conserve the particle number in each band are typically small~\cite{Groenewald2016}
and can be neglected.  In addition we assume a dependence only on the transferred
momentum, i.e.  $V_{lnmi}^{\blq\blk_{1}\blk_{2}}=V^{\blq}\d_{li}\d_{nm}$, and take the
standard 2D interaction $V^{\blq}=2\pi/\e(q+q_{c})$, where $\e$ accounts for the
dielectric screening of the surrounding environment and $q_{c}$ 
encodes the ground state
screening from the filled bands.  Typical values to describe optical excitation in a
monolayer transition metal dichalcogenide (TMD) around the K valley are $ \e_{g}=2$~eV,
$m=0.5m_{e}$ ($m_{e}$ being the electron mass) and $\e=10$ (e.g. sapphire substrate).  By
solving the Bethe-Salpeter equation at equilibrium with these parameters we find the
lowest energy exciton at $\e_{\mathrm{x}}\approx 1.9$~eV (i.e. binding energy of
$0.1$~eV).

\begin{figure}[tbp]
      \includegraphics[width=0.45\textwidth]{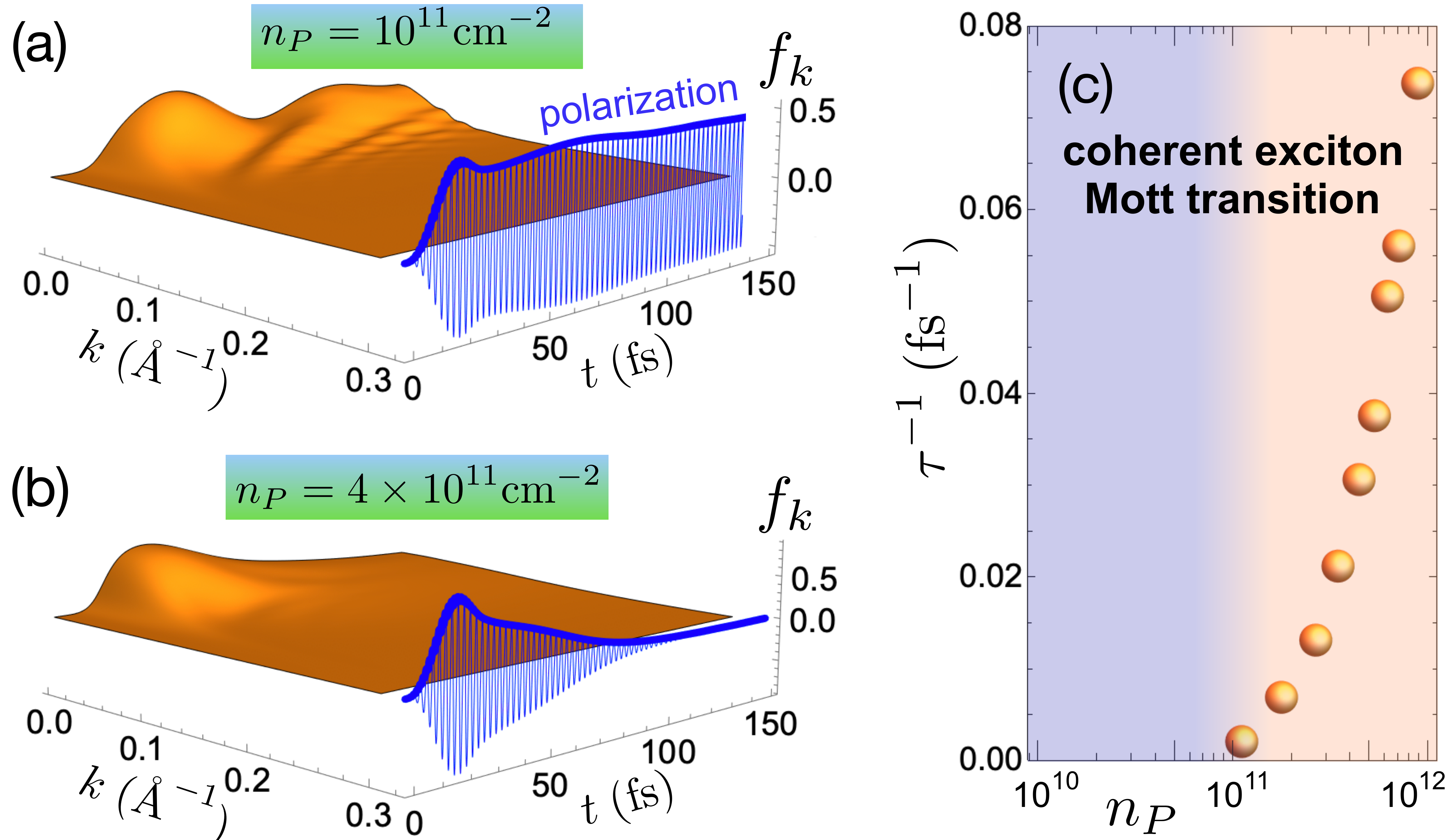}
   \caption{Real-time $GW$ evolution of the carrier distribution function $f_{k}(t)$ in a
     2D semiconductor for excitation density $n_{P}=10^{11}\mathrm{cm}^{-2}$ (a) and
     $n_{P}=4\times 10^{11}\mathrm{cm}^{-2}$ (b). The time-dependent polarization
     $p(t)$ (multiplied by a factor $2000$) is also shown as a blue curve.  In panel c we
     display the polarization inverse lifetime $\tau^{-1}$ as a function of $n_{P}$.  Here
     we used $q_{c}=0.02 ~ \AA^{-1}$.}
   \label{fig4}
\end{figure}

The system is excited with a laser pulse having the same enevelope $E(t)$ used for
graphene but with a resonant frequency $\w_{P}=\e_{\rm x}=1.9$~eV and duration
$T_{P}=25$~fs.  For simplicity we assume an isotropic excitation with momentum-independent
light-matter interaction $P_{\blk ij}(t)=\d_{i1}\d_{j2}M E(t)$. Accordingly the density
matrix and all observables depend only on the modulus $k$.  Also in this case the Rabi
frequency is varied in order to promote excitation densities in the range
$10^{10}-10^{12}$ carriers$/\mathrm{cm}^{2}$.  In Fig.\ref{fig4}(a) we show the evolution
of the momentum-resolved carrier distribution $f_{k}(t)=\r_{k 22}(t)$ for a low excited
density $n_{P}=\frac{4}{A}\sum_{\blk}\r_{k 22}(T_{P})= 10^{11}\mathrm{cm}^{-2}$ -- the
characteristic value $A=9\times 10^{-16}\mathrm{cm}^{2}$ has been used.  During pumping
excitons are prodominantly created and $f_{k} \propto |Y_{k}|^{2}$, where $Y_{k}$ is the exciton
wavefunction~\cite{PSMS.2019,PBS.2020,PhysRevB.103.L241404,man2021experimental,dong2021direct}.
We have recently shown that for small excited densities the coherent exciton superfluid is
not able to screen the Coulomb interaction~\cite{PhysRevB.102.085203}.  As a consequence
the $e$-$h$ attraction is not reduced and excitons survive for long time.
The superfluid phase is characterized by a macroscopic polarization $p(t)=\sum_{\blk}\r_{k
  12}(t)$ that oscillates monochromatically at the exciton frequency
$\e_{\mathrm{x}}$~\cite{PSMS.2019}, see the blue curve in Fig.\ref{fig4}(a).  In this
regime no relaxation occurs, and the carrier occupations $f_{k}(t)$ slowly attain the
values reached at the end of the pump.  The system can eventually thermalize only at later
times via electron-phonon scattering~\cite{PhysRevB.103.245103} (not considered in the
present work).

The scenario changes dramatically at higher excited densities.  In Fig.~\ref{fig4}b we see
that for $n_{P}=4\times 10^{11}\mathrm{cm}^{-2}$ the polarization damps in about $100$~fs,
and after few femtoseconds ($t \gtrsim 150$~fs) the occupations $f_{k}$ reach steady-state
values describing a Fermi-Dirac distribution at temperature $\sim 2000$~K (not shown),
consistently with recent data~\cite{li2020ultrafast}.  We have systematically studied the
lifetime of the polarization $p(t)$ by varying the excitation density.  In Fig.~\ref{fig4}
we plot the inverse of the time $\tau$ needed to reduce the amplitude of $p(t)$ by one
order of magnitude, as function of the carrier density $n_{P}$.  We see that no damping of
the polarization can be detected for $n_{P}\lesssim 10^{11}\mathrm{cm}^{-2}$, while
$\tau^{-1}$ grows very fast beyond this threshold.  The mechanism behind this behavior is
a screening cascade: (1) at sufficiently high excitation density the screening of the
excitonic superfluid is nonvanishing and the effective $e$-$h$ attraction is reduced (2)
excitons start dissociating in a plasma of quasi-free electrons in conduction band and
quasi-free holes in valence band (3) the $e$-$h$ plasma has a high screening efficiency
and the $e$-$h$ attraction gets drastically reduced~\cite{PhysRevB.103.L241404}.  This
self-sustained mechanism leads to a rapid melting of the superfluid state, signaled by a
decay of the polarization.  The thermalization occurs via scattering between incoherent
quasiparticles, and the occupations $f_{k}(t)$ relax towards a hot Fermi-Dirac
distribution.  We emphasize that this {\it coherent exciton Mott transition} is different
from the well-known excitonic Mott
transition~\cite{Brinkman1973,Mott1973,RICE1978,Mott1949} which refers to the breakdown of
a system of incoherent excitons~\cite{Steinhoff2017}.

In conclusion we have demonstrated the feasibility of real-time $GW$ simulations in 2D
materials via the generalization and practical implementation of the recently proposed
$GW$-ODE scheme~\cite{PhysRevLett.124.076601,PhysRevB.101.245101}.  The $GW$ approximation
gives easy access to so far neglected effects that we show are crucial for the
photo-excited many-electrons dynamics in graphene and 2D semiconductors.  Although the
method presented in this work is devoted to purely electronic processes, it can be 
complemented with electron-phonon scatterings without affecting the linear-time
scaling~\cite{PhysRevLett.127.036402}.  This opens new avenues for 
the {\it ab initio} description and understaning of ultrafast phenomena observed in time-resolved experiments.

 \vspace{1cm}
 
{\it Acknowledgements} We acknowledge funding from MIUR PRIN Grant No. 20173B72NB and from
INFN17-Nemesys project.  G.S. acknowledges Tor Vergata University for financial support
through the Mission Sustainability Project 2DUTOPI.
%

\end{document}